\documentclass[iop,apjl]{emulateapj}

\slugcomment{Received 2016 Januray 11; accepted 2016 March 15}
\shorttitle{On the lamppost model of accreting black holes}
\shortauthors{Nied{\'z}wiecki et al.}

\begin{document}

\title{On the lamppost model of accreting black holes}

\author{Andrzej Nied{\'z}wiecki\altaffilmark{1}, Andrzej A. Zdziarski\altaffilmark{2} and Micha{\l} Szanecki\altaffilmark{1}}
\altaffiltext{1}{\L \'od\'z University, Department of Physics, Pomorska 149/153, 90-236 {\L}{\'o}d{\'z}, Poland}
\altaffiltext{2}{Centrum Astronomiczne im.\ M. Kopernika, Bartycka 18, 00-716 Warszawa, Poland}

\begin{abstract}
We study the lamppost model, in which the X-ray source in accreting black-hole systems is located on the rotation axis close to the horizon. We point out a number of inconsistencies in the widely used lamppost model \texttt{relxilllp}, e.g., the neglect of the redshift of the photons emitted by the lamppost and directly observed. They appear to invalidate those model fitting results for which the source distances from the horizon are within several gravitational radii. Furthermore, if those results were correct, most of the photons produced in the lamppost would be trapped by the black hole, and the luminosity generated in the source as measured at infinity would be much larger than that observed. This appears to be in conflict with the observed smooth state transitions between the hard and soft states of X-ray binaries. The required increase of the accretion rate and the associated efficiency reduction present also a problem for AGNs. Then, those models imply the luminosity measured in the local frame to be much higher than that produced in the source and measured at infinity, due to the additional effects of time dilation and redshift, and the electron temperature to be significantly higher than that observed. We show that these conditions imply that the fitted sources would be out of the e$^\pm$ pair equilibrium. On the other hand, the above issues pose relatively minor problems for sources at large distances from the black hole, where \texttt{relxilllp} can still be used.
\end{abstract}

\section{Introduction}
\label{intro}

We consider the lamppost geometry \citep{mm96,mf04} as applied to X-ray emission of accreting black holes (BHs). In it, a point-like X-ray source is located on the BH rotation axis perpendicular to a surrounding flat disk. The BH has a dimensionless angular momentum, $a$, at which the horizon is at $r_\mathrm{H}=[1+(1-a^2)]^{1/2}$, and the innermost stable circular orbit (ISCO) is at $r_\mathrm{ISCO}(a)$ \citep{bardeen72}. The radial profile of the flux irradiating the disk is calculated using GR given the height of the source, $h$. The disk is truncated at an inner radius, $r_\mathrm{in}$. Hereafter $r_{\rm H}$, $r_\mathrm{ISCO}$, $h$ and $r_\mathrm{in}$ are expressed in units of the gravitational radius, $R_\mathrm{g}\equiv{GM/c^2}$, where $M$ is the BH mass. 

The resulting spectra in this geometry have been calculated in the \texttt{relxilllp} model \citep{garcia14a} in \textsc{xspec} \citep{arnaud96}. That model combines the \texttt{xillver} code \citep{gk10}, which uses the atomic data of \texttt{XSTAR} \citep{xstar} to calculate angle-dependent reflection spectra, with the relativistic blurring code \texttt{relconv} \citep{dauser10}. The lamppost model as implemented in \texttt{relxilllp} has been widely used to model X-ray emission from accreting BHs. Sources for which the lamp height was claimed to be within a few $R_\mathrm{g}$ from the horizon include the BH binaries Cyg X-1 \citep{parker15}, GX 339--4 \citep{furst15} and the Seyfert galaxies Mrk 335 \citep{parker14}, NGC 4151 \citep{keck15}. Here, we consider such sources.

We compare \texttt{relxilllp} with the code \texttt{reflkerr} of \cite{nz08} and \cite{nm10}. The latter assumes reflection to be neutral. Given that reflection in accreting systems is usually from ionized media, we do not perform spectral fitting with \texttt{reflkerr}, but concentrate on comparing the treatment of the relativistic effects in those two codes. 

\section{Model comparison}
\label{comparison}

In our comparison of \texttt{relxilllp} and \texttt{reflkerr}, we have found a number of differences. The main one concerns the treatment of the primary spectral component. The \texttt{relxilllp} model takes into account the reduction of the direct flux (emitted isotropically in the local frame) due to light bending (causing a fraction of the emitted photons to either cross the BH horizon or hit the disk). However, the spectrum of the emitted radiation, both that reaching the observer and that irradiating the disk, is not redshifted (which is also the case for the model using disk emissivity profiles, \texttt{relxill}). This causes the intrinsic cutoffs in the direct spectrum (usually considered to originate from Comptonization, e.g., \citealt{zg04}) to be substantially underestimated. In particular, the e-folding energy, $E_\mathrm{c}$, fitted using \texttt{relxilllp} has to be multiplied by $(1+z)$ (where $z$ is the redshift as seen at infinity) to get its intrinsic value. Also, the neglect of the redshift of the radiation irradiating the disk (which is different from that of the radiation directly received, and radius-dependent) strongly impacts the shape of the observed reflected component. 

The second important difference is in the amplitude of the reflected component in the exact lamppost geometry (i.e., setting $\texttt{fixReflFrac}=1$ in \texttt{relxilllp}). This property of \texttt{relxilllp} differs between the current (v0.4a) and earlier versions, and we note problems with both. In the earlier versions, the treatment of \citet{dauser14} is followed, in which the amplitude is calculated by setting the ratio of the number of photons incident on the disk, $n_\mathrm{irr}$, to those escaping to infinity, $n_\mathrm{esc}$, equal to the ratio of the 20--40 keV energy flux in the reflected component to the corresponding direct one. The $n_\mathrm{irr}/n_\mathrm{esc}$ ratio is calculated averaging over all directions. However, the effect of reflected photons crossing the BH horizon (due to light bending) is neglected, which results in an overestimate of the actual reflection fraction. Furthermore, while the direct photons emitted by an isotropic source on the symmetry axis arrive at infinity approximately isotropically, the reflected photons are strongly bent toward the equator. This causes the flux reflected at a low inclination angle, $i$, to be overestimated and the one at a high inclination, to be underestimated, which then affects assessments of $h$ based on the fitted reflection strength (e.g., \citealt{parker14}). The current version (v04.a) uses a reflection fraction parameter \citep{dauser16} taking into account relativistic transfer of reflected radiation from the disk. We find that the reflection amplitude of v0.4a roughly agrees with that of \texttt{reflkerr} for $h\la2$. However, at larger $h$, we find differences by a factor of several; at $h=100$, where relativistic effects are minor, \texttt{relxilllp} predicts the reflection flux lower by a factor of $\sim$6 for high $i$ than both \texttt{reflkerr} and the static-reflection \texttt{pexrav} model \citep{mz95}, which in turn agree well with each other.

The next difference is due to \texttt{relxilllp} not taking into account the disk irradiation by returning (due to light bending) reflected radiation. At $a \ga0.95$ and $r_\mathrm{in}=r_\mathrm{ISCO}$, including the 2nd-order reflection significantly increases the amplitude (up to a factor of $\sim$5 at $a=0.998$ and $h=1.3$) and changes the reflection spectrum.

\begin{figure}
\centering{\includegraphics[width=8.2cm]{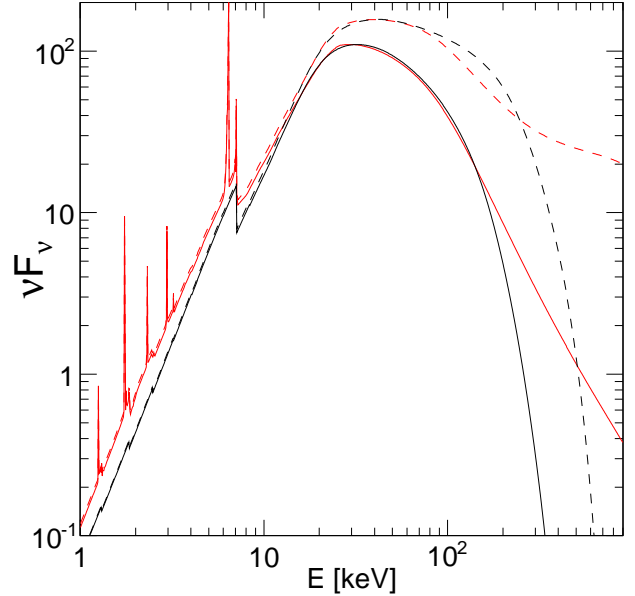}}
\caption{Comparison of the reflection spectra from \texttt{xillver} (red curves) and \texttt{pexrav} (black curves) for $i=18\degr$, the photon index of $\Gamma=1.75$, and $E_\mathrm{c}=150$ keV (solid curves) and $E_\mathrm{c}=1$ MeV (dashed curves). The reflecting medium is neutral for \texttt{pexrav} (and line emission is not included), while the ionization parameter is $\xi=1$ for \texttt{xillver}. The \texttt{xillver} spectra are normalized to match those of \texttt{pexrav} at 30--60 keV ($\texttt{refl\_frac}\simeq1.2$).}
\label{xillver_pexrav}
\end{figure}

Finally, the non-relativistic treatment of Compton scattering in \texttt{xillver} (and thus also in \texttt{relxilllp} and \texttt{relxilll}) results in the reflected flux at high energies calculated inaccurately. This is illustrated in Figure \ref{xillver_pexrav} (see also \citealt{bz16}). For $E_\mathrm{c}=150$\,keV, the spectrum of \texttt{xillver} becomes incorrect at $E\gtrsim150$ keV. The difference increases with increasing value of $E_\mathrm{c}$, and for $E_\mathrm{c}=1$\,MeV, \texttt{xillver} becomes inaccurate at $E\gtrsim60$ keV. In models with $r_\mathrm{in}=r_\mathrm{ISCO}$, $a\simeq1$ and low $h$, such as analyzed here, the bulk of reflected radiation is redshifted by $(1+z)\sim$2--3, and emission observed at $\ga$20 keV is then affected. This may significantly affect the accuracy of the determination of $E_\mathrm{c}$ using reflection fitting in the method of \citet{garcia15}. On the other hand, the differences seen at $\lesssim20$ keV appear to be due to different choice of abundances and atomic cross sections, with \texttt{xillver} being probably more accurate in that range (although its assumption of the irradiation at a constant $45\degr$ angle may contribute to these differences).

As the specific example of the source parameters we use the results of \citet{keck15} for NGC 4151. Those authors obtained $a=0.98$, $h=1.3$ for the lamppost geometry using \texttt{relxilllp}. Given that \texttt{relxilllp} neglects the redshift of the primary radiation, we compare its results with those of the exact treatment (of \texttt{reflkerr}) using both the nominal cutoff energy and its value $(1+z)$ times higher than $E_\mathrm{c}$ of \texttt{relxilllp}. For $a=0.98$, $h=1.3$, $z\simeq6$.
 
\begin{figure} 
\centerline{\includegraphics[width=8.2cm]{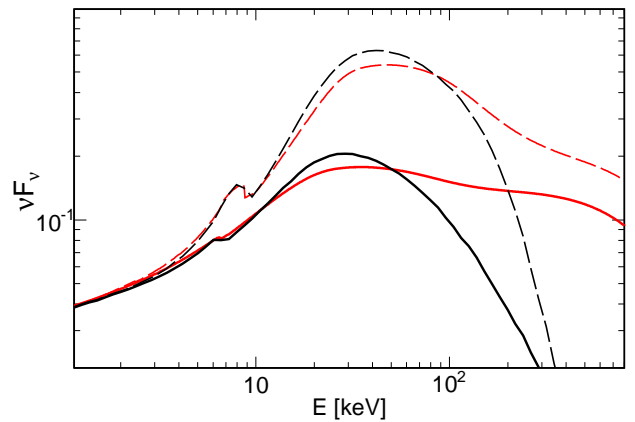}}
\caption{The observed spectra (direct emission+reflection) for the lamppost model with $h=1.3$, $a=0.98$, computed with \texttt{relxilllp} at $\texttt{fixReflFrac}=1$ (red curves) and with \texttt{reflkerr} (black curves). The heavy solid and thin dashed curves are for $i=18\degr$ and $76\degr$, respectively. The direct spectrum has $\Gamma=1.75$ and $E_\mathrm{c}=1$\,MeV. The spectra are normalized to the 1-keV flux.}
\label{h1_3} 
\end{figure}

\begin{figure} 
\centerline{\includegraphics[width=8.2cm]{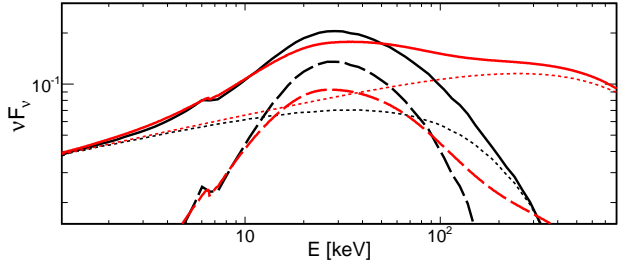}}
\caption{Spectral components for the low-inclination case of Figure \ref{h1_3} ($\Gamma=1.75$, $E_\mathrm{c}=1$ MeV, $h=1.3$, $a=0.98$, $i=18\degr$). The solid black curve gives the actual total spectrum (obtained with \texttt{reflkerr} including the 2nd-order reflection, same as the solid black curve in Figure \ref{h1_3}). The red solid curve is the result of \texttt{relxilllp} with $\texttt{fixReflFrac}=1$. The dotted and dashed curves give, in both cases, the direct and reflected components, respectively.}
\label{h1_3_components} 
\end{figure}

Figure \ref{h1_3} compares the results of \texttt{relxilllp} with the exact treatment for the above parameters for two viewing angles, and without correcting $E_\mathrm{c}$. We see major differences between \texttt{relxilllp} and the exact treatment for both angles. The properly applied redshift causes the observed high-energy cutoff to be 7 times lower than that of \texttt{relxilllp}. Figure \ref{h1_3_components} shows the reflected and direct components for the low-inclination case ($i=18\degr$) shown in Figure \ref{h1_3}. We see that both the direct and reflected components are overestimated by \texttt{relxilllp} at high energies.

\begin{figure} 
\centerline{\includegraphics[width=8.2cm]{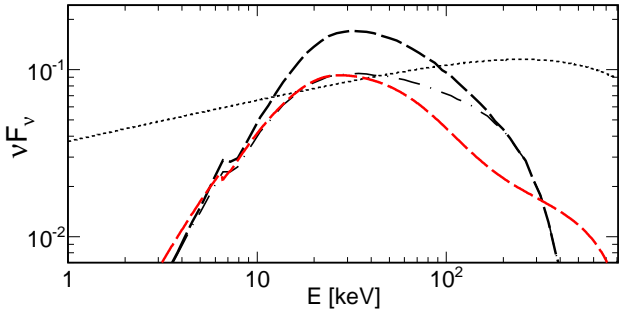}}
\caption{The spectral components for the same case as in Figure \ref{h1_3_components} ($i=18\degr$) but now with compensation of the neglect of redshift in \texttt{relxilllp} by using $E_\mathrm{c}$ in \texttt{reflkerr} higher by $(1+z)$. The observed $E_\mathrm{c}=1$\,MeV then corresponds to the intrinsic value of 7\,MeV. The black dotted curve shows the observed intrinsic spectra (identical for \texttt{reflkerr} and \texttt{relxilllp}) and the black and red dashed curves show the reflected components for \texttt{reflkerr} and \texttt{relxilllp} with $\texttt{fixReflFrac}=1$. The dot-dashed black curve shows the reflection of \texttt{reflkerr} without the 2nd-order component.}
\label{compensation} 
\end{figure}

We then perform the comparison attempting to compensate for the neglect of the redshift by using $E_\mathrm{c}(1+z)$ in \texttt{reflkerr} vs.\ $E_\mathrm{c}$ in \texttt{relxilllp}. The results are shown in Figure \ref{compensation}. We see that the reflected spectra, shown by the black and red dashed curves, still disagree. The cause for the 1st-order reflected spectra being lower in \texttt{relxilllp} than in the actual spectrum is the neglect of the radius-dependent redshift of the primary radiation incident on the disk in the former. That redshift seen from inner parts of the disc is $\sim$2, i.e., significantly lower than that of the directly observed radiation, $z\simeq6$. Thus, the compensation for the neglect of redshift in \texttt{relxilllp} still does not correspond to an actual physical situation. The 2nd-order reflection leads to further distortion of the reflected spectrum both at low and high energies, compare the black dashed and dot-dashed curves. 

\section{The luminosity and mass accretion rate}
\label{luminosity}

If the intrinsic emission of accreting BH sources occurs indeed very close to the horizon, we have to take into account a large fraction of the emitted photons crossing it due to light bending. In the case of NGC 4151, the result of \citet{keck15} of $h=1.3$, $a=0.98$, yields $n_\mathrm{esc}\simeq0.01$ (while $n_\mathrm{esc}=0.5$ for an isotropic source above a disk in the flat space-time). This implies that the source is strongly photon-advection dominated, with the actual (i.e., including photons trapped by the BH) luminosity as measured at infinity $\simeq$50 times larger than that observed. Since the observed X-ray luminosity of NGC 4151 is $\sim10^{-2}L_\mathrm{Edd}$ (where $L_\mathrm{Edd}$ is the Eddington luminosity), the implied accretion rate is $\dot{M}\sim10^{-2}L_\mathrm{Edd}/(2n_\mathrm{esc}\epsilon_\mathrm{accr}c^2)\gtrsim1.5L_\mathrm{Edd}/c^2$, where $\epsilon_\mathrm{accr}\lesssim0.3$ is the accretion efficiency neglecting advection. Then the overall radiative efficiency, $L/\dot{M}c^2$, is $\lesssim10^{-3}$. Such low radiative efficiencies cannot be common in AGNs, given their average accretion efficiency of $\gtrsim0.1$ \citep{soltan82,marconi04,silverman08,schulze15}. Also, if such high accretion rates are common in AGNs, the implied BH mass growth rates will be much faster than currently thought. 
 
In the case of the hard state of BH binaries, we consider here the result of \citet{parker15} for Cyg X-1. They give their best-fit results only graphically. From their fig.\ 7, the best fit is obtained at $a\simeq1$, $h\simeq1$. Their stated confidence limits are $a>0.97$, $h<1.56r_\mathrm{ISCO}$ (but note that there is no apparent physical requirement for $h\geq{r_\mathrm{ISCO}}$). However, $a$ and $h/r_\mathrm{ISCO}$ are strongly correlated in the fits, see their fig.\ 7, which implies that the maximum and minimum allowed $h/r_\mathrm{ISCO}$ is achieved at the maximum and minimum allowed $a$, respectively. Then, the constraint of $a>0.97$ implies $h\la1.7$. For those parameters, we find $n_\mathrm{esc}\leq0.058$, corresponding to a reduction of the accretion efficiency by $\ga$9 (and much more at their best fit).

This presents a major problem for our understanding of state transitions of BH binaries. The commonly accepted model for their soft spectral state (e.g., \citealt{done07}) is an optically thick accretion disk extending to the ISCO, at which the dissipation disappears (expressing the zero-stress inner boundary condition). For such disks, a half power is emitted within $\simeq33R_\mathrm{g}$ for $a=0$, and within $\simeq5R_\mathrm{g}$ for $a=1$. Thus, even for $a=1$ the photon-advection reduction of the radiative efficiency is minor. On the other hand, most of the intrinsically emitted luminosity in the hard state in the lamppost model goes below the BH horizon, which then implies a large luminosity jump at transitions between the hard and soft state. However, the observed luminosity changes are minor, by a factor of $\lesssim$2 \citep{frontera01,z02,z04,malzac06}. Thus, in the lamppost model, this constraint implies that the accretion rate in the hard state is higher than in the soft state. This is highly unlikely, and in disagreement with observations of transient BH sources, in which the hard-to-soft transition occurs during the initial outburst rise, and the reverse transition occurs during the decline.

\begin{figure} 
\centerline{\includegraphics[width=7.5cm]{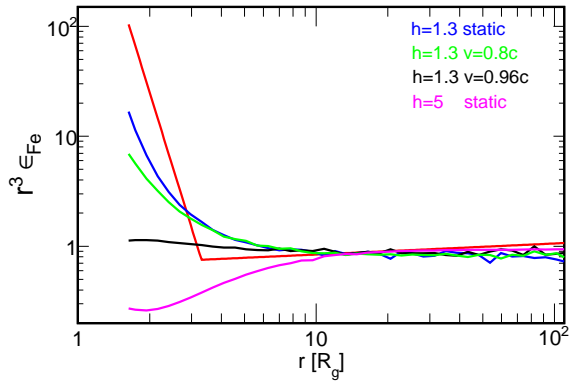}}
\caption{Radial profiles times $r^3$ for $a=0.98$ at $h=1.3$ for a static lamppost (blue curve) and those moving up with $v=0.8c$ (green curve) and $0.96c$ (black curve; here $n_{\rm esc}\simeq0.5$), and at $h=5$ for a static source (magenta curve). Also, we show the $r^{-10}$ profile below $r=3.3$ and $r^{-2.8}$ above it fitted to NGC 4151 by \citet{keck15}. We see its inner part is much steeper than even the static lamppost profile at $h=1.3$.}
\label{outflow} 
\end{figure}

We could reduce the effect of the photon trapping by the BH if the source moves relativistically away from the BH. However, this would also strongly reduce the relativistic broadening of the Fe line, making the irradiation profile close to $\propto{r^{-3}}$, and removing most of the steep part of the profile close to the BH, characteristic to the lamppost model. The reduction of the reflection effects due to an outflow in the coronal geometry was pointed out by \citet{beloborodov99}. This effect is shown in Figure \ref{outflow}, where, in contrast to the coronal case, we need a highly relativistic velocity for this mechanism to be effective at $h=1.3$. We also note that the static lamppost irradiation profile is still much shallower than $r^{-10}$ fitted by \citet{keck15} in their model of NGC 4151 with a broken power-law disk emissivity profile. Thus, that phenomenological profile cannot be reproduced even at $h=1.3$. We also note that the static profile for $h=3$ is already $\propto{r^{-3}}$ at all $r$ and even flatter for $h=5$, see Figure \ref{outflow}.

\section{Effects on \lowercase{e}$^\pm$ pair production}

The condition of equilibrium between production and annihilation of e$^\pm$ pairs can be expressed in terms of the electron temperature, $T_\mathrm{e}$, and the compactness parameter, e.g., \citet{svensson84}, 
\begin{equation}
\ell\equiv\frac{L_\mathrm{intr}\sigma_\mathrm{T}}{{d}R_\mathrm{g}m_\mathrm{e}c^2}=\frac{{4\pi}m_{\rm p}L_\mathrm{intr}}{{d}m_\mathrm{e}L_\mathrm{Edd}},
\label{compactness}
\end{equation}
where $d$ is the characteristic size in unit of $R_\mathrm{g}$, $L_\mathrm{intr}$ is the luminosity in the local frame, $\sigma_\mathrm{T}$ is the Thomson cross section, and $m_\mathrm{e}$ and $m_\mathrm{p}$ is the electron and proton mass, respectively. Calculations of pair equilibrium using spectra from thermal Comptonization were performed by \citet{z85,stern95}. Their constraints on pair balance at $kT_\mathrm{e}=0.1$, 1\,MeV were $\ell\lesssim10^2$--$10^3$, 0.1--1, respectively.

\citet{fabian15} proposed that pair production is important in AGNs and BH binaries, and the composition of their X-rays sources is dominated by pairs produced in photon-photon collisions. Their arguments were based on calculating $T_\mathrm{e}$ from the observed high-energy cutoffs and using the source luminosities as observed. They did then estimate a correction due to the gravitational redshift in an approximate way, and noted the existence of the luminosity correction (see below), which, however, was not taken into account. Their results remain approximately correct for sources with sizes $\gg{R_\mathrm{g}}$.

Here, we consider cases with low $h$ and use the fitted parameters of NGC 4151 and Cyg X-1. As noted above, $z\simeq6$ for the NGC 4151 fit, which would make the actual plasma temperature 7 times higher than that estimated from the observed cutoff. \citet{keck15} fixed $E_\mathrm{c}=1$\,MeV in their modeling. However, their data extend to only 80 keV, which does not allow a precise determination of $kT_\mathrm{e}$. Still, the lack of a visible cutoff constrains $kT_\mathrm{e}/(1+z)\gtrsim150$\,keV (assuming the observer's frame temperature is at least twice the maximum observed energy), and thus $kT_\mathrm{e}\gtrsim1$\,MeV. At this temperature, the results of \citet{stern95} require $\ell\lesssim0.1$--1. 

We use then the source luminosity in the local frame, $L_\mathrm{intr}$. Compared to the observed $L$, it is corrected not only for the photon trapping by the BH, relevant for a remote observer, but also for the time dilation and redshift (since we now consider the local conditions). Taking all those factors into account, we find that the intrinsic luminosity is higher by a factor $\simeq2\times10^3$ than the observed $L$. Thus, $L_\mathrm{intr}\gtrsim20L_\mathrm{Edd}$. As argued by \citet{dd16}, the size of the lamppost has to be substantially less than the height. Here, we assume $d\lesssim(h-r_\mathrm{H})/2$. From equation (\ref{compactness}) and for $h=1.3$, we thus find $\ell\gtrsim3\times10^6$. This places that model very deep in the region forbidden by pair equilibrium, at which the production rate is orders of magnitude higher than the annihilation rate. 

Similarly, the best-fit model ($a\simeq1$, $h\simeq1$) of Cyg X-1 by \citet{parker15} is very far from pair equilibrium. We consider thus the model with the largest allowed $h\simeq1.7$ and the lowest $a\simeq0.97$. The observed temperature is $kT_\mathrm{e}/(1+z)= 43\pm 1$\,keV, the observed luminosity is $0.025L_\mathrm{Edd}$, the fraction of escaping photons is $n_\mathrm{esc}<0.058$, and $(1+z)\gtrsim3$, which results in $L_\mathrm{intr}\simeq1.9L_\mathrm{Edd}$. We assume $d\lesssim(h-r_\mathrm{H})/2\simeq0.65$. This yields $\ell\gtrsim7\times10^4$. For $kT_\mathrm{e}\simeq130$\,keV, the curves in fig.\ 1 of \citet{stern95} yield $\ell\lesssim (1$--$3)\times10^2$, while fig.\ 3a of \citet{z85} yields $\ell\lesssim50$. Thus, the model is out of pair equilibrium in the allowed parameter range.

Therefore, pair production allows us to constrain the sizes and locations of the accreting X-ray sources. They have to have some minimum size to reduce the effects of photon trapping, redshift and time dilation sufficiently in order to obtain the intrinsic compactness and electron temperature satisfying the pair equilibrium condition. The strong gravity corrections on the equilibrium condition are essential for $h\la3$, at which $L_\mathrm{intr}$ is by a factor of $\ga10$ larger than the observed $L$. We also note that the corona has to be large enough if the seed photons for Compton scattering are from the surrounding disk \citep{dd16}.

On the other hand, we find no pairs in the X-ray sources in the hard state of GX 339--4 when we use the sizes estimated from fitting X-ray spectra by \citet{bz16} and from reverberation by \citet{demarco15}. They both found $r_\mathrm{in}\sim{d}\ga30$ at the highest observed $L/L_\mathrm{Edd}\simeq0.15$, implying $\ell\la10^2$. The temperature at bright states of GX 339--4 is $\sim$50\,keV \citep{wardzinski02}, which then implies the absence of pairs \citep{stern95}. For lower $L$, larger $r_{\rm in}$ were found, which yields even lower $\ell$, and thus still no pairs. 

\section{Conclusions}

We have pointed out a number of inconsistencies in the widely used lamppost model \texttt{relxilllp}. The neglect of the redshift of the source radiation is a major deficiency for $h\la5$, where $(1+z)\ga1.3$, and similarly in \texttt{relxill} models assuming steep irradiation profiles. The non-relativistic treatment of Compton scattering may then affect results of fitting data at $\ga20$\,keV with small $h$ or steep radial profiles due to the redshift of the observed reflection. Fits with \texttt{relxilllp} are usually done allowing for a free amplitude of the reflection component. In that case, the obtained reflection strength may be incorrectly interpreted physically, as earlier versions of \texttt{relxilllp} used an oversimplified definition of the reflection strength, and the current version still gives incorrect reflection fraction for most of the parameter space. Results of fitting models with $a\simeq0.998$ and $r_\mathrm{in}\simeq{r_\mathrm{ISCO}}$ need to be viewed with particular caution, as the 2nd-order reflection (neglected in \texttt{relxilllp}) is much stronger, and different in shape, than the 1st-order one for $h\la3$. The above inconsistencies appear to invalidate a number of results of fitting this model to hard-state spectra of accreting BHs in which X-ray source locations very close to the BH horizon were obtained. 

We have also considered consequences of those results for the accretion rates. Given that majority of the photons produced in the lamppost at $h\la3$ are trapped by the BH, the produced power as measured at infinity has to be much larger than the observed $L$. This requires a large increase of the accretion rate and a decrease of the accretion efficiency. This is then in conflict with the observed smooth state transitions in BH binaries between the hard and soft states, and with the observed average AGN accretion efficiency. 

The extreme locations of the modeled X-ray sources also imply that the luminosity measured in the local frame is much higher than even the produced power as measured at infinity, due to time dilation and redshift. The electron temperature is also significantly higher than that observed due to the redshift. This implies that such a source has the pair production rate much higher than the annihilation rate. This provides one more argument against the physical reality of models with $h\la3$.

Finally, we note that \texttt{relxilllp} and \texttt{relxill} can still be used to model weak-gravity effects in the energy range of $\la$80\,keV. The problems pointed out here result then in only minor corrections to fit results.

\acknowledgments
We thank Chris Reynolds, the referee, for valuable comments, and Chris Done for discussions on outflows. This research has been supported by the NCN grants DEC-2011/03/B/ST9/03459, 2012/04/M/ST9/00780, 2013/10/M/ST9/00729.

\end{document}